\RequirePackage{fix-cm}
\documentclass[smallextended]{svjour3}       
\usepackage{graphicx}
\usepackage{amsmath}
\usepackage{bm}
\allowdisplaybreaks
\journalname{Few-body Systems}
\begin{document}

\title{Investigating the Transverse Momentum Dependent Gluon Sivers Function
in  Quarkonium production at $pp$ colliders
}

\titlerunning{Gluon Sivers Function in Quarkonium Production}

\author{U. D'Alesio         \and
        F. Murgia         \and
        C. Pisano
}


\institute{U. D'Alesio \and C. Pisano \at
              Dipartimento di Fisica, Universit\`a di Cagliari, Cittadella Universitaria, Monserrato (CA), 09042, Italy \\
              \\ \email{umberto.dalesio@ca.infn.it,  cristian.pisano@unica.it}           
           \and
           U. D'Alesio \and F. Murgia \and C. Pisano \at
              Istituto Nazionale di Fisica Nucleare, Sezione di Cagliari, Cittadella Universitaria,
              Monserrato (CA), 09042, Italy \\
              \email{francesco.murgia@ca.infn.it}
}

\date{Received: date / Accepted: date}

\maketitle

\begin{abstract}
In this contribution, we will present a short overview of the transverse momentum dependent
(TMD) approach as a tool for studying the 3-dimensional structure of hadrons in high-energy
(un)polarized hadron collisions.
We will then summarize the present status of a running research programme that aims at constraining
the poorly known transverse momentum dependent gluon Sivers function, through
the study of single spin asymmetries in quarkonium (mainly $J/\psi$), pion, and $D$-meson production
in polarized proton-proton collisions at RHIC.
Finally, we will shortly discuss perspectives for this field of research,
emphasizing in particular its role in the physics programme of LHC in the fixed-target setup and NICA.

\end{abstract}

\section{Introduction}
\label{intro}
An in-depth knowledge of the nucleon structure in terms of its elementary constituents (quarks and gluons) and
of their internal orbital motion, as well as of the confinement mechanism and parton fragmentation processes is now more than ever
mandatory in order to fully understand strong interactions. This is important also in view of the intense efforts in the search for unambiguous signals of new physics in processes where hadrons are involved.
Understanding the fundamental mechanisms and properties of strong interactions is not only relevant for hadron spectroscopy, that has recently known a renaissance in the quarkonium and hybrids sector (not to mention the ever open issue of glueballs). Indeed, there are many aspects and facts in the dynamical behaviour of hadrons involved in high-energy collision processes that are still awaiting full interpretation and comprehension. In particular, in the last decades, it has become clear  that many interesting observables sensitive to the intrinsic parton motion inside hadrons survive at much larger energies and transverse momenta than expected.
It is by now clear that the usual ``collinear" perturbative QCD approach to (un)polarized high-energy hadronic processes is not able to fully account for these experimental findings. By ``collinear" QCD we mean the approach based on collinear factorization theorems, in which parton intrinsic motion is neglected as compared to the dominant longitudinal motion along the hadron direction. In this scheme, soft contributions are described by one-dimensional parton distribution and fragmentation functions (respectively PDFs and FFs), depending only on the parton collinear momentum fraction and, through perturbative evolution, on the energy scale involved.
In this contribution, we will mainly focus on the so-called Transverse Momentum Dependent (TMD) approach, where the degrees of freedom related to intrinsic parton motion, and their correlation with the spin of the particles involved, are explicitly taken into account. As we will discuss in section~\ref{sec:TMD}, this leads to a generalization of the usual one-dimensional PDFs and FFs into a larger class of transverse momentum dependent soft functions, named in general TMDs, mapping information on the 3-dimensional structure of hadrons.
In section~\ref{sec:GSF} we will present an application of this approach to the study of single spin asymmetries in quarkonium production in $pp$ collisions. Finally, in section~\ref{sec:con} we gather our conclusions.

\section{The Transverse Momentum Dependent approach}
\label{sec:TMD}
The role of intrinsic parton motion in high-energy hadronic processes has been investigated already at the formulation of the parton model by Feynman and his collaborators, see e.g.~\cite{Feynman:1978dt}. In the early eighties, following the advent of perturbative QCD, in a series of well-known papers Ellis, Furmanski and Petronzio studied the role of power-corrections, higher-twists and parton off-shellness within the formalism of perturbative QCD, with special attention to formal aspects related to intrinsic parton motion and Lorentz invariance~\cite{Ellis:1982cd}. Intrinsic transverse momentum effects were also tentatively taken into account in Drell-Yan processes in order to explain the lower end of the lepton-pair transverse momentum spectrum~\cite{Feynman:1978dt}.
However, all these pioneering studies considered only unpolarized processes, therefore missing the basic ingredient of what is now known as ``TMD approach", that is the possible correlations among the intrinsic motion of partons inside hadrons (or hadrons resulting from parton fragmentation) and their spin and polarization vectors.
Taking explicitly into account the spin and transverse momentum degrees of freedom enlarges the number of independent non-perturbative functions playing a role in polarized processes.
As an example, while in collinear pQCD we have three independent leading-twist~(LT) distribution functions for quarks inside a proton, the unpolarized $q(x)$, the longitudinally polarized (or helicity) $\Delta q(x)$ and the transversely polarized (or transversity) $\Delta_Tq(x)$ distributions, in the TMD approach we can have up to eight independent, LT parton distributions, according to the polarization state of the quark and the parent nucleon. A similar situation holds for gluon TMD PDFs inside the nucleon, and for hadron fragmentation functions: while in the collinear approach for (pseudo)scalar or unpolarized hadrons at leading twist there is only one independent quark FF, in the TMD approach, besides the unpolarized FF we can also have the so-called Collins FF, describing the fragmentation of a transversely polarized quark into an unpolarized hadron. Similarly, there  are eight leading-twist TMD FFs for a quark fragmenting into a spin-$1/2$ hadron (like e.g.~a $\Lambda$ hyperon). Beyond leading twist an even increasing number of possible independent TMD PDFs and FFs appears as compared to the collinear case, some of them being related by Lorentz invariance and QCD equations of motion.

It is important to stress that many of these new TMD functions by themselves vanish if integrated over intrinsic parton motion. However, their effects may survive in measurable spin and azimuthal asymmetries at hadronic level because of kinematical and polarization correlations among at least two of them (both in the distribution sector, like for Drell-Yan (DY) processes, or in the fragmentation sector, like in $e^+e^-\to h_1 h_2 + X$ processes, or one in the distribution and one in the fragmentation sectors, like in semi-inclusive deeply inelastic scattering (SIDIS)).

We will not discuss in detail here the complete class of leading-twist TMDs, focusing on the most relevant ones, from the historical and phenomenological point of view, both in the distribution and in the fragmentation sector. A full account on the TMD approach and TMD PDFs and FFs can be found in a series of dedicated reviews, see e.g.~\cite{DAlesio:2007bjf,Barone:2010zz,Anselmino:2020vlp}.

The (quark and gluon) TMD Sivers distribution function~\cite{Sivers:1989cc} describes the asymmetry in the azimuthal distribution (around the direction of motion of the parent hadron) of non-collinear partons inside a high-energy, transversely polarized proton. This asymmetry is in principle allowed by the preferential direction introduced by the transverse polarization of the proton with respect to the plane containing the proton-parton system.
The Sivers distribution plays a role in several spin and azimuthal asymmetries measured in polarized proton-proton collisions, in SIDIS and in DY processes.
In fact, it was originally introduced in order to explain the large single spin asymmetries (SSAs) measured in inclusive pion production in polarized proton-proton collisions. Sivers azimuthal asymmetries are by now known to be definitely non-zero in SIDIS processes. Since gluons do not couple directly to the virtual photon, observation of the gluon Sivers distribution in SIDIS requires considering processes like heavy-quark pair, double quarkonium or quarkonium plus jet production.

The Sivers function is naively T-odd and chiral-even, therefore can couple to other chiral-even TMD functions, like the unpolarized TMD PDFs or FFs, in order to produce observable single spin or azimuthal asymmetries.

Conversely, the Boer-Mulders TMD quark distribution function~\cite{Boer:1997nt} describes the asymmetry in the azimuthal distribution of transversely polarized non-collinear quarks inside a high-energy, unpolarized proton (a similar distribution can be defined also for linearly polarized gluons). It was originally employed~\cite{Boer:1999mm} to explain the dilepton azimuthal asymmetries (in the center of mass system of the pair) in Drell-Yan processes in the lower end of the dilepton transverse momentum spectrum. Unfortunately, in SIDIS processes the measurement of azimuthal asymmetries generated by the Boer-Mulders TMD PDF is hindered by the mixing with other possible TMD effects, both at leading (Cahn effect) and subleading twist. As a consequence, phenomenological information is not presently at the same level as for the Sivers distribution.
Like the Sivers function, the Boer-Mulders function is naively T-odd, but it is also chiral-odd (due to the transverse polarization of the quark). Therefore, it can give place to observable asymmetries only when convoluted with another chiral-odd TMD function, either in the distribution or in the fragmentation sector, depending on the specific process under consideration.

Unpolarized TMD parton distributions can play a relevant role for azimuthal distributions in unpolarized SIDIS cross sections (the so-called Cahn effect). Their role is also essential for the description of the transverse momentum spectrum of produced hadrons in SIDIS and of their multiplicities. Being all these observables cross sections (and not ratios of cross sections like the asymmetries), their study is much more delicate both from the experimental and theoretical side.

In the fragmentation sector, for unpolarized hadrons, besides the unpolarized TMD fragmentation function, a preminent role is played by the Collins function~\cite{Collins:1992kk}
(a similar function for the fragmentation of linearly polarized gluons can be defined).
Like the Boer-Mulders function in the distribution sector, the Collins FF is T-odd and chiral-odd, therefore appears in spin asymmetries only in connection with another chiral-odd TMD function  (for example, convoluted with the TMD quark transversity distribution in SIDIS and other single inclusive processes, or with a second Collins function in hadron pair production in $e^+e^-$ annihilations).
Also in the fragmentation sector, for spin-$1/2$ hadrons there are eight leading-twist TMD FFs, in clear analogy with the distribution case. Apart from the unpolarized and the Collins FF, a relevant role is played by the so-called \emph{polarizing} FF, a sort of mirror analogue of the Sivers function, describing the azimuthal asymmetry in the distribution of transversely polarized spin-$1/2$ hadrons produced in the fragmentation of a high-energy unpolarized quark. This function can be responsible, in a TMD approach, for the large transverse polarization of $\Lambda$ hyperons inclusively produced in unpolarized proton-proton and proton-nucleus collisions, measured since the late 70s and still waiting a clear explanation in perturbative QCD and, more recently, in $e^+e^-$ annihilations (see, e.g., \cite{DAlesio:2020wjq}).

The TMD approach is a mature field of research counting on over twenty years of intense theoretical and experimental activity, witnessed by a vaste literature. Many interesting and relevant results have been achieved in due time and many others are under active investigation concerning several aspects of the approach. TMD physics is now facing a new, more sophisticated stage, progressing more and more from the initial semi-qualitative level to a very detailed quantitative one, comparable in accuracy to the collinear pQCD analysis of PDFs and FFs, at least for some processes.
It is not possible in this short overview to fully account for this activity, for which we address the reader to dedicated reviews~\cite{DAlesio:2007bjf,Barone:2010zz,Anselmino:2020vlp}. In the sequel, we briefly outline some of the more interesting facets of the TMD approach currently under active investigation.\\
{\it Factorization, universality and process dependence}.
As it is well known, the predictivity power of pQCD is based on the validity of factorization theorems and the consequent universality of the soft functions (PDFs and FFs) describing the nonperturbative phase, combined with perturbatively calculable hard parton scattering processes.
While in the early stages of TMD physics simplified models were mostly adopted, aiming at first qualitative results, nowadays TMD factorization theorems have been proven for several of the most interesting processes, in particular for SIDIS and Drell-Yan processes and for hadron pair production in $e^+e^-$ collisions. Somehow ironically (but understandably given the complex nature of the process) no TMD factorization theorem has been proven yet for inclusive hadron production in hadronic collisions, where the first (and still larger in size) asymmetries were measured, giving a decisive boost to the TMD research field. Rather, there are indications that factorization might be broken e.g.~for inclusive two-particle production. However, such processes are so relevant from the phenomenological point of view, giving valuable (and at present unique) information on the process dependence, large-$x$ behaviour and flavour separation of TMDs, that it is of the utmost importance to keep studying them, performing detailed phenomenological analyses.
A relevant role in this respect is played by the so-called generalized parton model (GPM)~\cite{Anselmino:2005sh}, and its more recent extension, named the colour-gauge invariant GPM(CGI-GPM)~\cite{Gamberg:2010tj,DAlesio:2011kkm}. Applications of these phenomenological TMD approaches to the study of the gluon Sivers function (GSF) in proton-proton collisions will be the subject of the second part of this contribution. In the GPM factorization is taken as a reasonable assumption, extending the collinear QCD improved parton model with the inclusion of polarization and intrinsic motion effects, and TMDs are assumed to be universal; conversely, in the CGI-GPM they may be process dependent due to initial- and final-state interactions.\\
{\it Gauge invariance}.
PDFs and FFs are obtained by proper Dirac projections of the corresponding hadronic correlators.
In order to guarantee gauge invariance, appropriate Wilson lines, connecting the position of the fields contained in the correlators, must be inserted. In the collinear case, it is possible to work in the axial gauge, where these gauge links reduce to unity operators and can be neglected. The situation is different in the TMD approach, where transverse momentum effects are taken into account explicitly. In this case, one cannot operate simply with collinear gauge links
and work in the axial gauge. Rapidity divergences appear and transverse gauge links are required. As a result, they do not anymore reduce to unity operators and their effects need to be taken into account.
Due to this, TMDs may acquire explicit process dependence, calculable perturbatively, and modified universality behaviours.
The most known example is the predicted change of sign of the Sivers function measured in Drell-Yan processes as compared to that extracted from SIDIS processes. This is considered as a fundamental test of the TMD approach and of QCD in general. Available data on Sivers single spin asymmetries from DY processes at RHIC~\cite{Adamczyk:2015gyk} and COMPASS at CERN~\cite{Aghasyan:2017jop} are not precise enough yet to test this prediction unambiguously, although giving some indications in favour of it.\\
{\it TMD evolution}.
Dependence on the hard scale of TMDs is another important issue, based on factorization and the derivation of appropriate TMD evolution equations. In the last years, significant developments have been reached. Starting from a parametrization of the TMD function at a low, nonperturbative scale, and working in the conjugate impact-parameter space, complete evolution equations, including Sudakov resummation factors, have been formulated~\cite{Collins:2011zzd}, in some cases up to next-to-next-to-leading order (NNLO) and next-to-next-to-next-to-leading logarithms (N$^3$LL) accuracy, see e.g.~\cite{Scimemi:2019cmh}.
However, the phenomenological application of the formal scheme is complicated by the scarce information available on the nonperturbative input describing the large impact parameter behaviour. Different choices seem to lead to significantly different outcomes in the evolution with scale, which are not well under control.
As a matter of fact, while evolution effects are surely crucial for cross section and multiplicity studies, at present there is no significant experimental evidence of them in spin and azimuthal asymmetries, where they can partially cancel.\\
%
{\it Flavour and $k_\perp$ dependence}.
Information on the flavour and explicit transverse momentum dependence of TMDs is currently limited and not well constrained by available experimental data. This is mainly due to the fact that most of the information comes from processes like DY, SIDIS and $e^+e^-$ annihilations, in kinematic regimes where weak interactions, which would give a much stronger flavour separation, are negligible (see however~\cite{Adamczyk:2015gyk}). Moreover, results on spin and azimuthal asymmetries are often given as a function of Bjorken $x$, $Q^2$, and the fraction of energy of the observed final hadron $z$, integrated over transverse momenta.
Many phenomenological analyses and fit procedures assume flavour independence for the transverse behaviour of TMDs, and simplified Gaussian, or Gaussian-like shapes. In fact, more detailed choices would introduce additional parameters that could be strongly correlated and largely unconstrained by present data. From this point of view, we have to wait for more precise, multidimensional experimental input.\\
%
{\it Twist-three approach}.
Let us finally just mention a relevant alternative approach to spin and azimuthal asymmetries.
The collinear twist-three approach is an extension of the usual collinear pQCD formalism with the inclusion of higher-twist contributions to hadronic correlators via quark-gluon-quark and three-gluon correlators.
This formalism, for which factorization theorems have been proven, was originally developed at the beginning of the 90s for the explanation of the large SSAs measured in single inclusive pion production at large transverse momentum in polarized proton-proton collisions~\cite{Qiu:1991pp,Kouvaris:2006zy}. The twist-three approach is indeed applicable when there is only one (large) energy scale in the process, namely the transverse momentum of the observed particle or jet. Therefore, it is also applicable to SIDIS but again in the region where the transverse momentum of the produced hadron is large and comparable to the $Q^2$ of the process, a region therefore complementary to that covered by the TMD approach. For SIDIS, it has been shown that the two approaches match in the overlapping region.
One has also to take into account that the connection between SIDIS and proton-proton collisions, necessary for testing the process dependence and universality of the new collinear twist-three distributions, is an indirect one: The higher-twist contributions in the hadronic correlators playing a role in proton-proton collisions are in fact related to moments (in transverse momentum space) of the corresponding TMDs in SIDIS processes.


\section{Gluon Sivers function in polarized $pp$ collisions and the (colour gauge invariant) generalized parton model}
\label{sec:GSF}

In the second part of this contribution we concentrate on an ongoing analysis that aims at constraining the poorly known gluon Sivers function through a phenomenological analysis of inclusive single particle production in polarized proton-proton collisions in kinematical configurations dominated by the gluon-fusion partonic process.
Since, as discussed in the previous section, factorization has not been proven for this class of processes, we will adopt the generalized parton model, that assumes factorization as a phenomenological starting Ansatz and extends the collinear QCD-improved parton model via the inclusion of spin and parton intrinsic motion effects, adopting the helicity formalim. We will also consider its colour gauge invariant extension, that takes into account the role of proper Wilson lines in the hadronic correlators and the corresponding initial-state (ISIs) and final-state (FSIs) interactions in a leading-order expansion of the gauge link, that is via the exchange of a single eikonalized gluon between the particles involved in the hard scattering processes and the remnants of the polarized proton beam. In the case of quarkonium production, in conjunction with the TMD approach we will use the nonrelativistic QCD (NRQCD) formalism for describing the formation of the final observed quarkonium state from the hadronization of the heavy $Q\bar{Q}$ pair created in the hard scattering process, through the multiple emission of soft gluons.

NRQCD was originally formulated~\cite{Bodwin:1994jh} to cope with both phenomenological and formal problems of the colour-singlet (CS) model, in which the $Q\bar{Q}$ pair created in the hard partonic process has already the quantum numbers of the observed quarkonium state. In the CS approach the soft nonperturbative input of the model is entirely encoded in the value of the bound-state radial wave function at the origin, in the full nonrelativistic limit. For $J/\psi$, $\Upsilon$ states, it can be directly inferred from the experimental dilepton decay widths, and simplifies
in ratios of cross sections, like e.g.~for spin and azimuthal asymmetries. At leading order in the strong coupling constant $\alpha_S$ the CS model was unable to explain (largely underestimating) the unpolarized cross sections for inclusive quarkonium production in $pp$ collisions. From a more formal point of view, it suffers from uncanceled infrared divergences in the decays of $P$-wave states.
NRQCD is an effective theory incorporating in a consistent way a double perturbative expansion, in $\alpha_S$ and in the relative velocity $v$ between the heavy quark and antiquark forming the quarkonium state. According to the specific quarkonium state and the kinematical configuration considered, one has to carefully order in relevance the contributions proportional to $\alpha_S^n v^m$ and truncate the perturbative expansion to some appropriate order. In NRQCD, the $Q\bar{Q}$ pair can be formed in the hard scattering process in any allowed $^{2S+1}L_J^{(c)}$ state, acquiring in a second step the correct quantum numbers of the observed quarkonium state through multiple emission of soft gluons. Here $S$, $L$, $J$ are respectively the spin, orbital and total angular momenta of the $Q\bar{Q}$ pair and $c=1,8$ is its colour (singlet or octet) state. In what follows, for the case of $J/\psi$ production, we will take into account the $^1S_0^{(8)}$, $^3S_1^{(1,8)}$ and $^3P_J^{(8)}$ ($J=0,1,2$) states.
Each of these possible contributions comes with an associated soft, non perturbative but in principle universal factor, known as long-distance matrix element (LDME), that must be fixed by fitting combinations of experimental results.
NRQCD is quite successful in explaining a wealth of data on inclusive quarkonium production in proton-proton and lepton-proton collisions, see e.g.~\cite{Brambilla:2010cs,Lansberg:2019adr}. However, several open points are still under debate, concerning in particular the confirmation of the universality of the soft LDMEs and the agreement with data on the spin alignment of $J/\psi$ and $\Upsilon$ states.

NRQCD has been and is mostly employed for the inclusive production of quarkonium states with large transverse momentum, in the context of collinear pQCD. The project summarized in this contribution aims at explaining spin and azimuthal asymmetries in quarkonium production in the low edge of the quarkonium transverse momentum spectrum, that is in the regime of validity of the TMD approach. The simultaneous use of the TMD and NRQCD approaches is surely a delicate task. However, we believe that the richness of experimental information available at present or in the near future in planned experimental setups, justifies efforts in this direction and offers a unique, additional tool for a deeper understanding and refinements of both approaches.

As we mentioned in the previous sections, the sizable transverse single spin asymmetries measured in single inclusive (charged and neutral) pion and kaon production in polarized proton-proton collisions at moderate hadron transverse momentum and large Feynman $x$ variable triggered the interest for early TMD approaches to spin and azimuthal asymmetries. Since large positive values of $x_F$ correspond (in a reference system with the $+\hat z$ axis along the direction of the polarized proton) to large values of light-cone momentum fractions for the struck parton inside the polarized proton, these data were particularly useful for testing the (valence) quark TMDs.
Since then, these pion SSA data were confirmed by the BRAHMS, PHENIX and STAR experiments at RHIC, at much larger center-of-mass (cm) energies~\cite{Aschenauer:2015ndk}.
Sivers and Collins effects have been separately confirmed also by the measurement of the corresponding azimuthal asymmetries in SIDIS in the valence region. All of these processes, however, give little or no information on the gluon Sivers function.
At leading order gluons decouple from all processes, SIDIS, DY and $e^+e^-$ annihilations, that are the pillars of TMD factorization. As a consequence, information on the GSF is very poor and limited. In order to investigate gluon TMDs one has to consider processes where the role of gluons in the hard scattering is enhanced and dominant above that of quarks. In this respect, inclusive particle production at low transverse momentum in the central rapidity region is the most favourable enviroment. Higgs and quarkonium production in gluon-fusion dominated proton-proton collisions are the most relevant examples, but a variety of other processes have been considered.
Among them we mention: inclusive jet, pion and photon production in proton-proton collisions at mid-rapidities, heavy-quark pair and quarkonium production, quarkonium plus jet or double jet production in proton-proton and lepton-proton collisions.

Experimental data on transverse single-spin asymmetries $A_N$ for some of the processes listed above, involving the gluon Sivers function, are available from the PHENIX and STAR Collaborations at RHIC. Although in some cases limited in statistics and relatively scarce in number, some of them can be already very powerful in partially constrain this function, as we will see in the sequel. Let us briefly summarize the available data of interest here:\\
1) Few data points for $A_N(p^\uparrow p\to J/\psi + X)$ at $\sqrt{s}=200$ GeV over the rapidity ranges $0.12 < |y| < 2.2$  and $|y| < 0.35$ for transverse momenta up to 6 GeV~\cite{Adare:2010bd,Aidala:2018gmp}.
Unpolarized cross section data for $J/\psi$ production in $p+p$ collisions in the central rapidity region are also available, both at $\sqrt{s} = 200$ GeV for $p_T$ up to 8.5 GeV~\cite{Adare:2009js} and at $\sqrt{s} = 500$ GeV for $p_T$ up to 20 GeV~\cite{Adam:2019mrg}.\\
2) $A_N$ in single inclusive jet~\cite{Adamczyk:2012qj} and neutral pion~\cite{Adare:2013ekj} production at $\sqrt{s} = 200$ GeV in the central rapidity region and for transverse momentum, $p_T$, approximately in the range $1 - 10$ GeV and $10 - 30$ GeV respectively for pions and jets.
In particular, data on neutral pion SSA are very precise for $p_T \sim 1 - 5$ GeV and can be therefore very effective in constraining the GSF in the region of validity of the TMD approach dominated by gluon-fusion subprocesses.
Data on jet production extends to larger $p_T$ but with lower statistics and precision.\\
3) $A_N$ for muons from open heavy-flavour decays in polarized $p+p$ collisions at $\sqrt{s}=200$ GeV for $-0.1 < x_F < 0.1$ and $p_T$ in the range $1-5$ GeV~\cite{Aidala:2017pum}.

In the sequel we will discuss how these results could represent a powerful tool for constraining the GSF. We will then give estimates for the Sivers asymmetries in processes and kinematical configurations of interest for future facilities, like
the proposed polarized fixed target setup at LHC, that will be crucial for detailed studies on the GSF and TMDs in general.
\subsection{Single spin asymmetry $A_N(p^\uparrow p \to J/\psi + X$) in the (CGI)GPM approach}
\label{sec:GSF-form}
The transverse single spin asymmetry for single-inclusive $J/\psi$ production in polarized $pp$ collisions can be defined as
\begin{equation}
A_N \equiv  \frac{{\rm d} \sigma^\uparrow - {\rm d} \sigma^\downarrow}{{\rm d} \sigma^\uparrow + {\rm d} \sigma^\downarrow} \equiv \frac{ {\rm d}\Delta\sigma}{ 2 {\rm d}\sigma}\,,
\label{eq:AN}
\end{equation}
where ${\rm d}\sigma^{\uparrow,\downarrow}$ is the invariant differential cross section $E_{\psi}
{\rm d}^3\sigma^{\uparrow,\downarrow}/{\rm d}^3\bm{P}_{\psi}$, and $\uparrow, \downarrow$ refers to the polarization of one of the colliding protons in the direction perpendicular to the production plane.
More precisely, we fix the $pp$ cm reference frame so that the $+\hat z$ axis is along the direction of the polarized proton beam, the $\hat x$-$\hat z$ plane coincides with the $J/\psi$ production plane, and the $\uparrow, \downarrow$ directions are respectively along the $\pm \hat y$ axis.
The numerator of the asymmetry, ${\rm d}\Delta\sigma$, receives its dominant contribution from the (quark and gluon, in principle) Sivers function that, as we have seen, describes the azimuthal asymmetry in the distribution of an unpolarized parton $a$ inside a transversely polarized proton $p$:
\begin{eqnarray}
\Delta \hat f_{a/p^\uparrow}(x_a, \bm{k}_{\perp a})  & = &
\hat f_{a/p^\uparrow}(x_a, \bm{k}_{\perp a}) - \hat f_{a/p^\downarrow}(x_a, \bm{k}_{\perp a}) =
 \\
\label{eq:defsiv}
\qquad &= & \Delta^N f_{a/p^\uparrow}(x_a, k_{\perp a}) \, \cos\phi_a \equiv
 -2 \, \frac{k_{\perp a}}{M_p} \, f_{1T}^{\perp a} (x_a, k_{\perp a}) \, \cos\phi_a \,,\nonumber
\end{eqnarray}
where in the second line the Torino-Cagliari notation, $\Delta^N f_{a/p^\uparrow}(x_a, k_{\perp a})$, and the Amsterdam one, $ f_{1T}^{\perp a} (x_a, k_{\perp a})$ have been used for the Sivers function~\cite{Bacchetta:2004jz}, and $\phi_a$ is the azimuthal angle of $\bm{k}_{\perp a}$. For a generic direction of the proton transverse-polarization vector, specified by the azimuthal angle $\phi_S$ (in the above defined reference frame $\phi_S = \pi/2$) the azimuthal dependence of the Sivers function is given by the factor $\sin(\phi_S-\phi_{a}$). Notice also that, somehow loosely, commonly the name Sivers function indicates both the full distribution including the azimuthal dependence and the one depending only on $k_{\perp a} = |\bm{k}_{\perp a}|$.

Assuming factorization, the cross sections ${\rm d}\sigma^{\uparrow,\downarrow}$ are written as usual as convolution integrals among the hard scattering cross sections for all allowed partonic processes and the corresponding soft TMD parton distributions. In the case of quarkonium production, the  NRQCD LDMEs are also included.

More specifically, for the $p^\uparrow p\to J/\psi + X$ process, in NRQCD at partonic level we have to consider two classes of contributions: a) $2\to 1$ processes, namely $g+g\to J/\psi$ and $q+\bar q\to J/\psi$; notice that these processes do not contribute to quarkonium colour-singlet states, due to colour imbalance; moreover, since in this case at leading order the quarkonium state acquires its transverse momentum, $P_T$, only from the intrinsic motion of the initial partons, $2\to 1$ processes can be relevant only in the low-$P_T$ region.
b) $2\to 2$ processes, that is $g+g\to J/\psi + g$, $g+q(\bar q)\to J/\psi + q(\bar q)$, $q+\bar q\to J/\psi +g$;
again, in the colour-singlet model, only the gluon-fusion channel contributes, while in the NRQCD approach also the quark Sivers function can play a role, associated with quarkonium colour-octet states. As said, for $J/\psi$ production we will take into account the contributions from the $^3S_1^{(1,8)}$, $^1S_0^{(8)}$ and $^3P_J^{(8)}$ $Q\bar Q$ states.

As already mentioned, we will consider both the generalized parton model and its colour gauge invariant extension, the CGI-GPM. The main difference among the two approaches is that in the GPM all TMD functions are as a working hypothesis considered universal, with the aim of severely scrutinize this Ansatz by comparing as many different processes as possible where the same TMDs play a role. On the contrary, in the CGI-GPM initial- and final-state interactions are taken into account at leading order via the exchange of a single eikonalized gluon among partons active in the hard processes and the remnants of the polarized proton. As a result, TMDs can be process dependent, but this dependence can be reabsorbed, in a perturbatively calculable way, into modified partonic hard cross sections. The other fundamental difference with respect to the GPM is that in its gauge invariant version there can be more than one type of certain TMDs. For example, in the CGI-GPM we can have two independent gluon Sivers functions, according to the different ways to perform colour recombination when the additional eikonalized gluon is involved. These so-called $f$- and $d$-type GSFs can have different properties and be subject to different general constraints, see e.g.~\cite{Boer:2015vso}.

\begin{figure}[t]
\begin{center}
\includegraphics[width=0.95\textwidth]{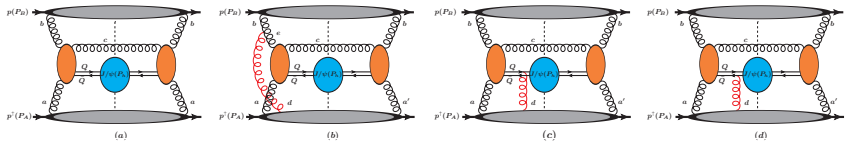}
\end{center}
\caption{Diagrams for the dominant gluon fusion contribution to the process $p^\uparrow p \to J/\psi + X$ in the GPM (a) and in the CGI-GPM approaches with inclusion, at leading order, of additional effects from initial-state (b) and final-state ((c) and (d)) interactions. FSIs are effective only when the $J/\psi$ is produced in a colour-octet state. Notice that there are analogous diagrams for other $2\to 2$ subprocesses as well as for the $2\to 1$ channels, like $g+g\to J/\psi$.
The scattering amplitudes for the underlying partonic reaction, $g+g\to J/\psi +g$, are represented by the central blobs, while the upper and lower ones describe the soft proton~$\to$~gluon transitions.
{\it Reprinted figure with permission from Ref.~\cite{DAlesio:2020eqo}, DOI:~https://doi.org/10.1103/PhysRevD.102.094011. Copyright (2020) by the American Physical Society}.}
\label{fig:ppjpsi}
\end{figure}

As an example, in Fig.~\ref{fig:ppjpsi} we show the Feynman diagrams for the process $p^\uparrow p \to J/\psi + X$ for the gluon-fusion contribution, in the GPM, Fig.~\ref{fig:ppjpsi}a, and in the CGI-GPM, including initial (Fig.~\ref{fig:ppjpsi}b) and final (Figs.~\ref{fig:ppjpsi}c,d) state interactions. The grey upper and lower blobs represent the proton $\to$ gluon + remnants soft  transitions; the sky-blue middle blob corresponds to the formation of the $J/\psi$ state from the $Q\bar Q$ pair;
the orange middle blobs correspond to the hard scattering process $gg\to Q\bar Q[^{2S+1}L_J^{(c)}] + g$, while the wavy red lines indicate the additional exchanged eikonalized gluon. Notice that FSIs vanish for colour-singlet states and also for the sum of diagrams where the additional eikonalized gluon attaches to the final unobserved gluon involved in the hard scattering process (for this reason, this case is not shown in Fig.~\ref{fig:ppjpsi}).

In relatively less complicated processes, like SIDIS or Drell-Yan, where there are only ISIs or FSIs at work, process dependence leads to overall factors and to generalized or modified universality properties for some TMDs, like the famous change of sign of the quark Sivers function between SIDIS and DY cases. For hadron production in $pp$ collisions, things are much more complicate: both ISIs and FSIs play a role and the number of partonic channels is typically larger.
However, it is possible to see that, at least for single partonic channels, or groups of Feynman diagrams for a given partonic channel, initial and final state interactions give rise to perturbatively calculable factors that can be collected and absorbed inside modified hard partonic cross sections. In summary, this means that in the CGI-GPM any product ${\cal M}{\cal M}^*$ of helicity amplitudes for the Feynman diagrams contributing to the partonic channels can be related to the corresponding one in the GPM unpolarized cross section as follows:
\begin{equation}
|{\cal M}^{\rm Inc}|^2 = \frac{C^{\rm Inc}}{C_U}\,|{\cal M}^U|^2 = \frac{C_I+C_F}{C_U}\,|{\cal M}^U|^2\,,
\label{eq:inc-unp}
\end{equation}
where the colour factor $C_U$ and the helicity amplitudes ${\cal M}^U$ refer to the unpolarized GPM process, while $C^{\rm Inc} = C_I+C_F$ and ${\cal M}^{\rm Inc}$ are the corresponding quantities for the CGI-GPM including initial ($I$) and final ($F$) state interactions. See~\cite{DAlesio:2017rzj,DAlesio:2018rnv,DAlesio:2019gnu,DAlesio:2020eqo} for more details and for tables collecting the colour factors for all Feynman diagrams contributing to the partonic processes involved in $J/\psi$ production.

Let us now summarize the expressions for the numerator, ${\rm d}\sigma^\uparrow - {\rm d}\sigma^\downarrow \equiv {\rm d}\Delta\sigma$, and the denominator,  ${\rm d}\sigma^\uparrow + {\rm d}\sigma^\downarrow \equiv 2{\rm d}\sigma^{\rm unp}$, of $A_N(p^\uparrow p\to J/\psi + X)$, starting with the CGI-GPM case:
\begin{eqnarray}
\label{eq:A2d1}
&& {\rm d}\Delta\sigma^{\mathrm{CGI-GPM}}_{2\to 1}
= \frac{2\pi}{x_a x_b s^2}\int d^2{\bm k}_{\perp a}d^2{\bm k}_{\perp b} \,\delta^2 ({\bm k}_{\perp a} + {\bm k}_{\perp b} - \bm{P}_T) \nonumber\\
& \times &  \Big(- \frac{k_{\perp a}}{M_p}\Big) \cos\phi_a
\Big\{\sum_{q} \Big[ f_{1T}^{\perp q} (x_a, k_{\perp a}) \,f_{\bar q/p}(x_b, k_{\perp b})\, |\mathcal{M}^{\mathrm{Inc}}_{q\bar q\to J/\psi}|^2\Big] \nonumber \\
& + & \sum_{m=f,d} f_{1T}^{\perp g(m)} (x_a, k_{\perp a})\, f_{g/p}(x_b, k_{\perp b}) \,|\mathcal{M}^{\mathrm{Inc}(m)}_{g g\to J/\psi}|^2 \Big\}\,,
\end{eqnarray}
\begin{eqnarray}
&& {\rm d}\Delta\sigma^{\mathrm{CGI-GPM}}_{2\to 2}
 = \frac{1}{(2\pi)^2}\frac{1}{2s} \int \frac{dx_a}{x_a} \frac{dx_b}{x_b}\,
d^2{\bm k}_{\perp a}\,d^2{\bm k}_{\perp b} \, \delta(\hat{s}+\hat{t}+\hat{u}-M^2_\psi)  \nonumber\\
& & \mbox{} \times \Big(-\frac{k_{\perp a}}{M_p}\Big) \cos\phi_a
\Big\{\sum_{q} \Big[ f_{1T}^{\perp q} (x_a, k_{\perp a}) \,\Big( f_{\bar q/p}(x_b, k_{\perp b})\, |\mathcal{M}^{\mathrm{Inc}}_{q\bar q\to J/\psi +g}|^2 \nonumber \\
&& \qquad\qquad +\, f_{g/p}(x_b, k_{\perp b})\, |\mathcal{M}^{\mathrm{Inc}}_{qg\to J/\psi+q}|^2\Big) \Big] \nonumber\\
&& \mbox{} + \sum_{m=f,d}f_{1T}^{\perp g(m)} (x_a, k_{\perp a}) \Big( \sum_q f_{q/p}(x_b, k_{\perp b})\, |\mathcal{M}^{\mathrm{Inc} (m)}_{gq\to J/\psi+q}|^2 \nonumber \\
&& \qquad\qquad\, + f_{g/p}(x_b, k_{\perp b})\,|\mathcal{M}^{\mathrm{Inc}(m)}_{g g\to J/\psi+g}|^2\Big) \Big\}\,.
\label{eq:N2to2}
\end{eqnarray}

The notation in these equations should be self-explicative, see however~\cite{DAlesio:2017rzj,DAlesio:2018rnv,DAlesio:2019gnu,DAlesio:2020eqo} for more details.
We only notice that, as already discussed, in the CGI-GPM we have to take into account two independent gluon Sivers functions, named $d$-type and $f$-type, with a terminology that recalls their origin in the two different ways to neutralize colour.
For $2\to 1$ processes, like in DY, the values of the light-cone momentum fractions are fixed and, neglecting higher-order corrections in $(k_\perp/\sqrt{s})$, are given by $x_{a,b}= \sqrt{(\bm{P}_T^2+M_\psi^2)/s}\,\exp{(\pm y)}$.

%
The simpler results for the GPM approach, without inclusion of ISIs and FSIs, can be obtained from Eqs.~(\ref{eq:A2d1}),~(\ref{eq:N2to2}), by taking only one single, universal term for the gluon Sivers function and replacing all ${\cal M}^{\rm Inc}$'s with the corresponding ${\cal M}^{U}$ amplitudes.
The expression of the denominators of the SSAs, neglecting higher-order initial- and final-state interactions, is the same in the GPM and CGI-GPM, and can be written as twice the unpolarized cross section:
\begin{eqnarray}
\label{2d1bis}
E_\psi\frac{d^3\sigma^{2\to 1}}{d^3{\bm P}_\psi} &=&
\sum_{a,b}\frac{\pi}{x_a x_b s^2}\int d^2{\bm k}_{\perp a}d^2{\bm k}_{\perp b}
f_{a/p}(x_a, k_{\perp a})f_{b/p}(x_b, k_{\perp b})\nonumber \\
&& \mbox{} \times \delta^2 ({\bm k}_{\perp a} + {\bm k}_{\perp b} - \bm{P}_T) |\mathcal{M}^U_{ab\rightarrow J/\psi }|^2\,,
\end{eqnarray}
\begin{eqnarray}
\label{d2}
E_\psi\frac{d^3\sigma^{2\to 2}}{d^3{\bm P}_\psi} &=&
\frac{1}{2(2\pi)^2}\frac{1}{2s}\sum_{a,b,c}\int \frac{dx_a}{x_a} \frac{dx_b}{x_b}\,  d^2{\bm k}_{\perp
a}d^2{\bm k}_{\perp b} f_{a/p}(x_a,k_{\perp a})f_{b/p}(x_b,k_{\perp b}) \nonumber \\
&& \mbox{} \times \delta(\hat{s}+\hat{t}+\hat{u}-M^2_\psi)
|\mathcal{M}^U_{ab\rightarrow J/\psi + c}|^2\,.
\end{eqnarray}

To go further in using these expressions, we need to fix the functional forms of the unknown TMD GSF and its unpolarized counterparts. We consider simple parameterizations for TMD distributions, factorizing the collinear and transverse momentum dependences.
More precisely, we take for the unpolarized TMD distributions (the dependence on the evolution scale is understood and not shown explicitly):
\begin{equation}
f_{a/p}(x_a,k_{\perp a}) =  f_{a/p}(x_a)\,\frac{e^{-k_{\perp a}^2/\langle k_{\perp a}^2 \rangle}}{\pi \langle k_{\perp a}^2\rangle}\,,
\label{eq:tmd-unp}
\end{equation}
where $f_{a/p}(x_a)$ is the corresponding ``collinear" PDF and we take $\langle k_{\perp q}^2 \rangle = 0.25$ GeV$^2$ and $\langle k_{\perp g}^2 \rangle = 1$ GeV$^2$ from fits to SIDIS unpolarized cross sections (for quarks)~\cite{Anselmino:2005nn} and to $J/\psi$ production in $pp$ collisions (for gluons)~\cite{DAlesio:2017rzj,DAlesio:2019gnu}. The Sivers function is parametrized as
\begin{equation}
\label{eq:siv-par}
\Delta^N\! f_{a/p^\uparrow}(x_a,k_{\perp a}) =   \frac{\sqrt{2e}}{\pi}   \,2\, {\cal N}_a(x_a)\, f_{a/p}(x_a) \, \sqrt{\frac{1-\rho_a}{\rho_a}}\,k_{\perp a}\, \frac{e^{-k_{\perp a}^2/ \rho_a\langle k_{\perp a}^2 \rangle}}
{\langle k_{\perp a}^2 \rangle^{3/2}}\,,
\end{equation}
with $0 < \rho_a < 1$,
\begin{equation}
{\cal N}_a(x_a) = N_a x_a^{\alpha_a}(1-x_a)^{\beta_a}\,
\frac{(\alpha_a+\beta_a)^{(\alpha_a+\beta_a)}}
{\alpha_a^{\alpha_a}\beta_a^{\beta_a}}\,,
\label{eq:nq-coll}
\end{equation}
and $|N_a| \leq 1$, so that the model-independent, natural positivity bound is satisfied for any values of $x_a$ and $k_{\perp a}$. The parameters $N_a$, $\alpha_a$, $\beta_a$ and $\rho_a$ have been fixed for valence quarks by fitting azimuthal Sivers asymmetries in SIDIS processes~\cite{Anselmino:2008sga}, while for the gluon Sivers function they are still subject of investigation. In the sequel we will first consider a scenario where the GSF is saturated to its maximized value allowed by the positivity bound, in order to study its potential role in $J/\psi$ production. For this, we will take $N_g =1$, $\alpha_g = \beta_g = 0$, that is ${\cal N}_g(x)=1$, and $\rho_g = 2/3$. We will also briefly mention a second scenario where, making use of available data on $A_N(p^\uparrow p\to \pi^0,\ D\to\mu^\pm + X)$ some significant constraints on the overall size of the GSFs can be imposed.
{}For the collinear, unpolarized PDFs, $f_a(x_a)$, we adopt the CTEQL1 set~\cite{Pumplin:2002vw}, taking $M_T=\sqrt{M_\psi^2+\bm{P}_T^2}$ as factorization scale, and DGLAP evolution. The other major ingredient that remains to be fixed is the set of long-distance matrix elements weighting the contribution of the different $^{2S+1}L_J^{(n)}$ $Q\bar Q$ states taken into account in our study. It is important to remind here that according to NRQCD factorization the LDMEs should be universal. In fact, one of the major problems faced by NRQCD at present is that there are sizable discrepancies among sets of LDMEs extracted from different phenomenological analyses.
\begin{figure}[t]
\begin{center}
\includegraphics[width=0.40\textwidth]{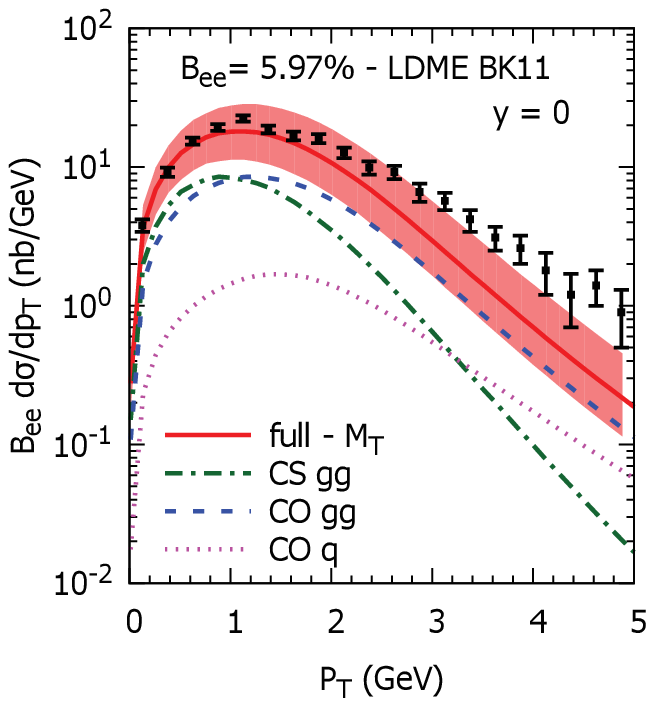}
\includegraphics[width=0.40\textwidth]{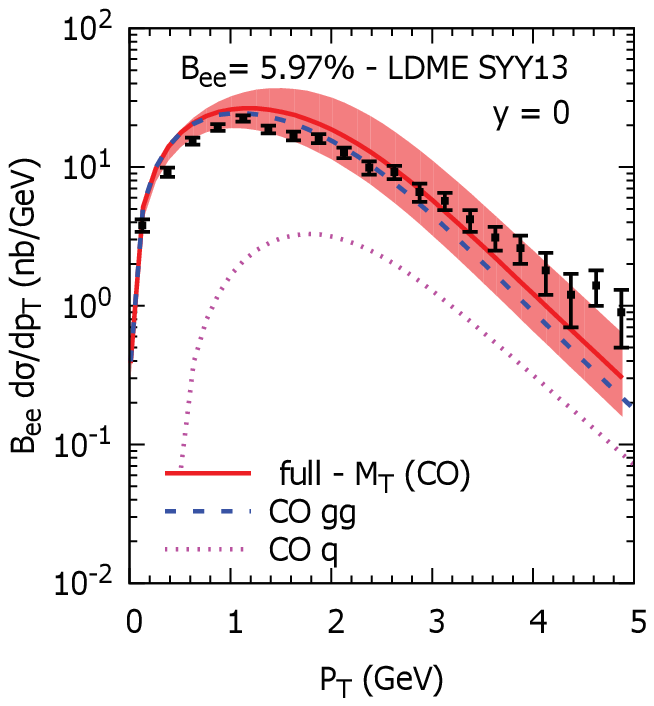}
\caption{Unpolarized cross section for the process $p p\to J/\psi + X$, as a function of $P_T$ at $\sqrt s=200$ GeV and $y=0$, within the GPM+NRQCD approach (red solid line) compared with PHENIX data~\cite{Adare:2009js}, obtained with the BK11 LDME set~\cite{Butenschoen:2011yh} (left panel) and with the SYY13~\cite{Sun:2012vc} one (right panel). The corresponding bands are obtained by varying the factorization scale $\mu$ from $M_T/2$ to $2M_T$. Separate contributions to the full cross section for $\mu=M_T$ are also shown: CS state (green dot-dashed line) and CO states for the $gg$ (blue dashed lines), $gq$, $qg$ and $q\bar q$ (magenta dotted lines) initiated subprocesses.}
\label{fig1:unpol}
\end{center}
\end{figure}
Moreover, most of the studies on quarkonium production in $pp$ collisions concentrate on the large-$P_T$ region of the quarkonium transverse momentum spectrum, typically $P_T > 5$ GeV. In our analysis, which combines the TMD and NRQCD approaches, we are rather interested in the low-$P_T$ range, where the TMD approach applies and NRQCD should be still reliable given the large mass of the produced quarkonium.
Therefore, we will employ two LDME sets extracted by considering also $J/\psi$ production data in the range $P_T <5$ GeV, the  Butenschoen and Kniehl (BK11) set~\cite{Butenschoen:2011yh} and the Sun, Yuan  and Yuan (SYY13) one~\cite{Sun:2012vc}.
These two sets differ significantly, for example the SYY13 set does not include at all the (small in this region) CS contribution, and other CO LDMEs differ in size and sign as compared to the BK11 set, leading to different mixtures of the various $^{2S+1}L_J^{(n)}$ contributions.
In the sequel we will adopt both LDME sets, in order to test the dependence of our results on this choice.
Although we are mainly interested in spin effects and single spin asymmetries, it is anyway important to check if our leading-order GPM + NRQCD approach reproduces in a reasonable way available data on unpolarized cross sections, at least in the small-$P_T$ range of interest here.
That this is indeed the case is shown in Fig.~\ref{fig1:unpol}, where our results are compared to PHENIX data~\cite{Adare:2009js}, both for the BK11 (left panel) and the SYY13 (right panel) LDME sets. We show the total result (red solid lines) with the corresponding uncertainty bands related to the choice of the factorization scale $\mu$ ($M_T/2 < \mu < 2 M_T$) and the partial contributions (at $\mu = M_T$) of the gluon-fusion CS (green dot-dashed line), and the quark-initiated (magenta dotted lines) and gluon-fusion (blue dashed lines) CO processes.
As said before, the SYY13 set neglects the colour-singlet contribution.

\begin{figure}[t]
\begin{center}
\includegraphics[width=0.42\textwidth]{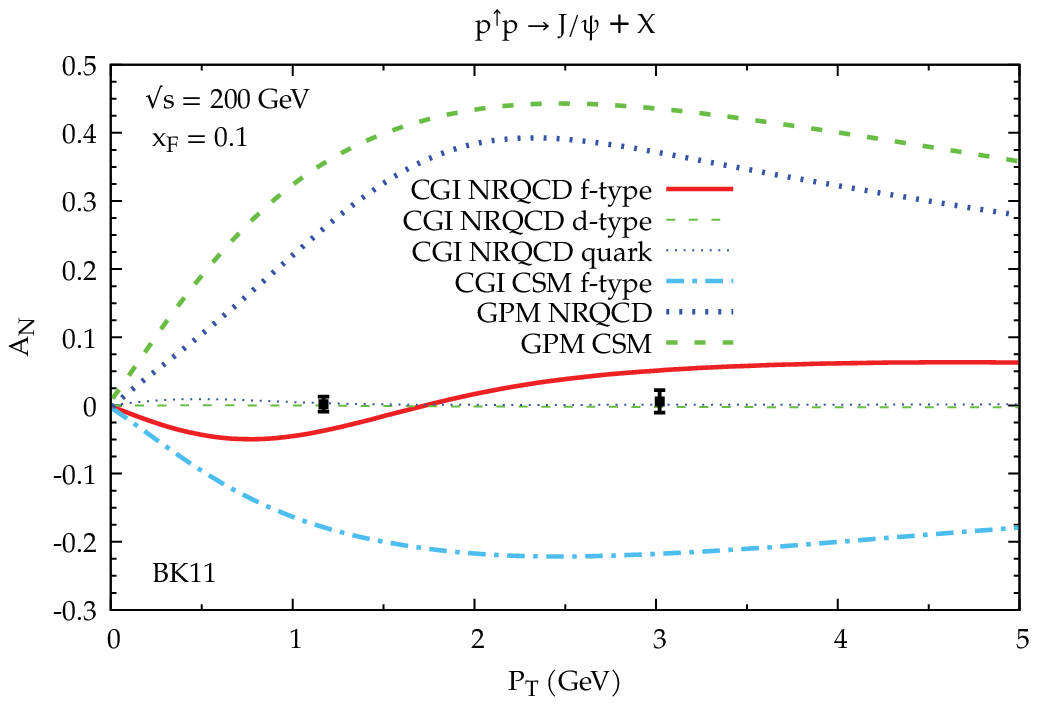}
\includegraphics[width=0.42\textwidth]{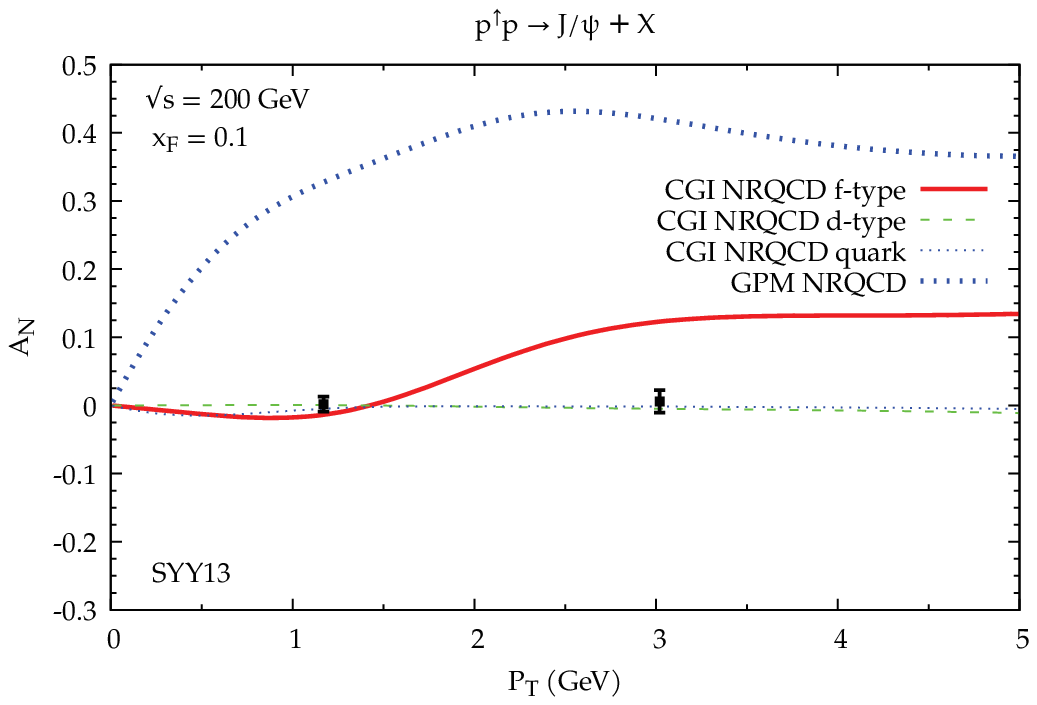}
\end{center}
\caption{Maximized contributions to $A_N(p^\uparrow p\to J/\psi + X)$ at $\sqrt s=200$ GeV and $x_F=0.1$, as a function of $P_T$, adopting the CGI-GPM and GPM approaches, within the CS model and NRQCD, for the BK11 (left panel) and the SYY13 (right panel) LDME sets. Data are taken from \cite{Aidala:2018gmp}.
{\it Reprinted figure with permission from Ref.~\cite{DAlesio:2020eqo}, DOI:~https://doi.org/10.1103/PhysRevD.102.094011. Copyright (2020) by the American Physical Society}.}
\label{fig:allxF01}
\end{figure}

\begin{figure}[t]
\begin{center}
\includegraphics[width=0.42\textwidth]{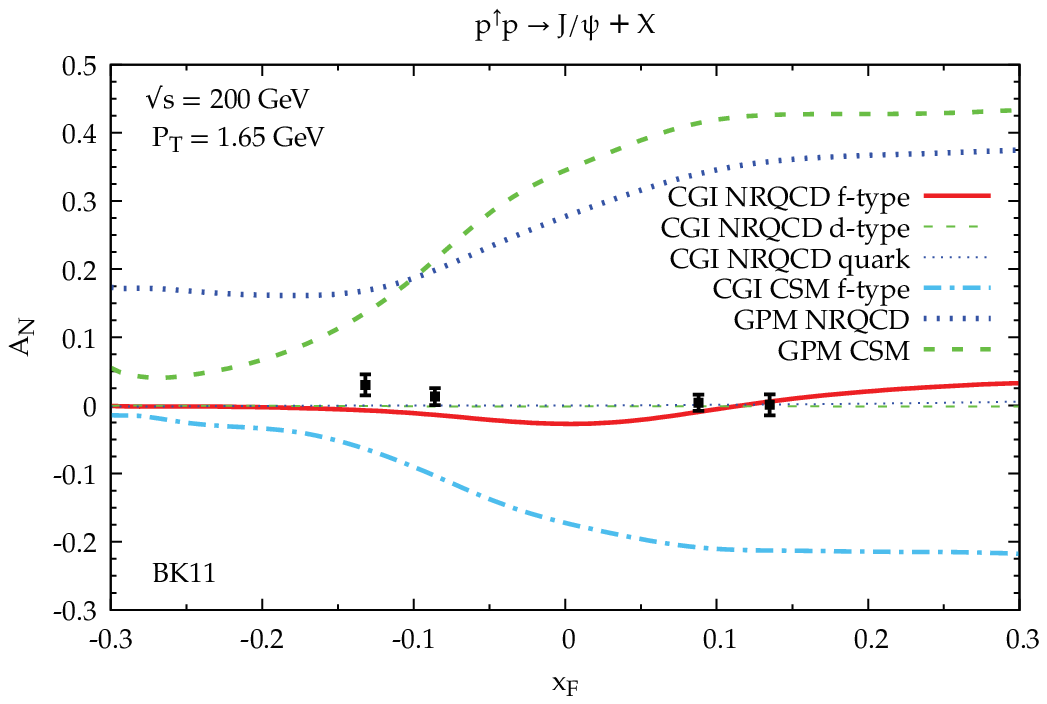}
\includegraphics[width=0.42\textwidth]{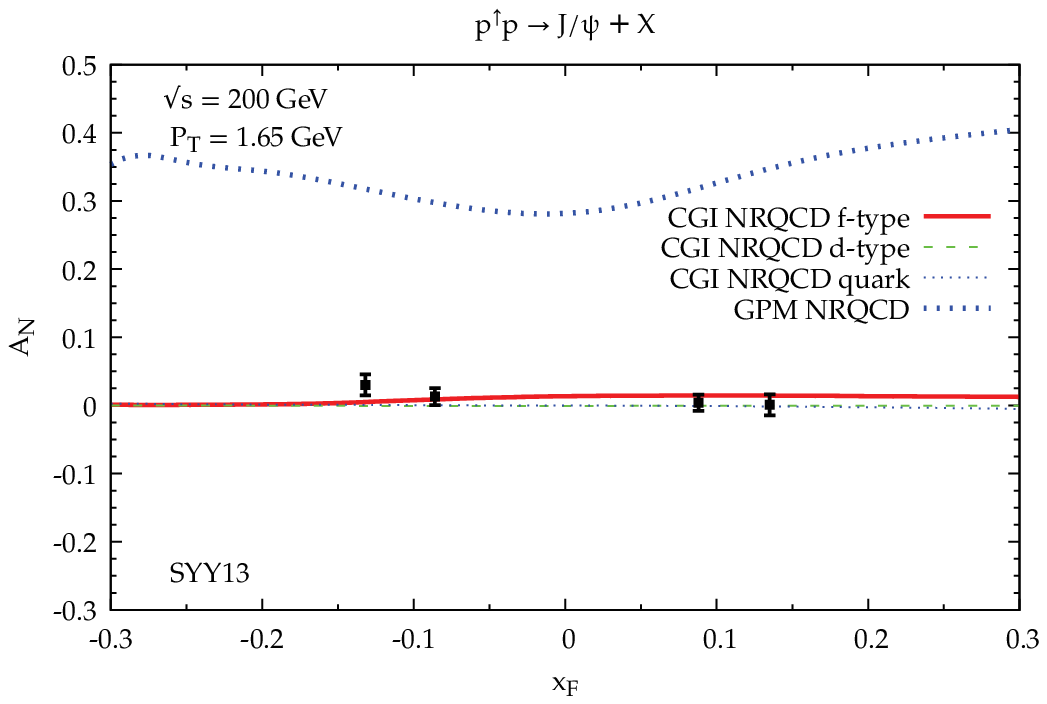}
\end{center}
\caption{Same as in Fig.~\ref{fig:allxF01}  but at fixed $P_T= 1.65$ GeV as a function of $x_F$.
{\it Reprinted figure with permission from Ref.~\cite{DAlesio:2020eqo}, DOI:~https://doi.org/10.1103/PhysRevD.102.094011. Copyright (2020) by the American Physical Society}.}
\label{fig:allpt165}
\end{figure}

Concerning the SSA $A_N(p^\uparrow p\to J/\psi + X)$, only few experimental data, almost compatible with zero, are available from the PHENIX Collaboration at RHIC~\cite{Aidala:2018gmp}. Let us now see wether and to what extent these data can help in constraining the GSF.
In Fig.~\ref{fig:allxF01} we compare the maximized contributions to $A_N(p^\uparrow p\to J/\psi + X)$ with PHENIX data at $\sqrt{s}=200$ GeV and $x_F=0.1$, as a function of $P_T$, adopting different combinations of the approaches considered: GPM-CS, CGI-CS, GPM-NRQCD and CGI-NRQCD (with separate quark and $f$- and $d$-type gluon contributions), again for the BK11 (left panel) and the SYY13 (right panel) LDME sets.
Several comments are in order: 1) The largest values are obtained in the GPM approach, both in the CS (thick green dashed line) and NRQCD (thick blue dotted lines) cases, the last one implying a moderate reduction in size as compared to the first. It is evident that in this approach the single, assumed universal GSF could be strongly constrained, as compared to its positivity bound, even by the few data available. 2) In the CGI-CS case (cyan dot-dashed line), where only the $f$-type GSF contributes, we find a change of sign and a reduction in size by a factor of 2, as compared to the GPM-CS case, since ISIs introduce process-dependent coefficients, absorbed into the hard scattering terms; also this case could be strongly constrained by PHENIX data. 3) As for the CGI-NRQCD case, the quark (thin blue dotted lines) and gluon $d$-type (thin green dashed lines) contributions are almost negligible, the bulk of the asymmetry being imputable to the $f$-type GSF (red solid lines). From these results we can make few additional comments: First, the asymmetry shows a clear oscillatory behaviour. This is basically due to subtle cancellations related to the dynamics of the different hard scattering contributions, as weighted by the Sivers azimuthal phase in the convolution integrals (see~\cite{DAlesio:2020eqo} for more details).
For the same reason, the size of $A_N$ depends crucially on the values of $P_T$ considered. The constraining power of the available data, as far as concerns the GSF, is much less effective as compared to the GPM case, however there is still some room left, in particular for the SYY13 LDME set.
%
Although the available data are scarce, their precision could be already sufficient for testing the oscillatory behaviour of the CGI-GPM results against the constant sign of the GPM ones and, at least for the largest $P_T$ data, put some valuable constraint on the size of the $f$-type GSF.
Since CS contributions are neglected in the SYY13 LDME set, the corresponding curves do not appear in the right panel of Fig.~\ref{fig:allxF01}. On the other hand, these contributions are independent of the particular set of LDMEs considered, since a unique LDME appears and cancels out between the numerator and the denominator of the SSA. Therefore, the same curves appearing in the left panel hold also for the right one.
In Fig.~\ref{fig:allxF01} we have considered only results at fixed $x_F = 0.1$. Two mirror experimental points are available also at negative $x_F = - 0.1$. In general, investigating the negative Feynman $x$ regime is interesting both for looking at changes in the interplay among the different partonic contributions, as weighted by the corresponding TMD distributions and by the Sivers azimuthal phase appearing in the convolution integrals of the numerator and the denominator of the spin asymmetry. Therefore, in Fig.~\ref{fig:allpt165} we report the same maximized results presented in Fig.~\ref{fig:allxF01}, but this time as a function of $x_F$ at fixed $P_T = 1.65$ GeV, corresponding indicatively to the average $P_T$ of the PHENIX data.
Qualitatively, the behaviour of the various models and of their different contributions is similar to that shown in Fig.~\ref{fig:allxF01}: The GPM CS and NRQCD results can be potentially large and have the same, definite sign, with the NRQCD case being moderately suppressed as compared to the CS one (notice however the reversed situation in the negative $x_F$ region); The CGI-CS result is reversed in sign and reduced by a factor of 2 as compared to the GPM-CS, but still potentially large; The quark and $d$-type gluon Sivers contributions in the CGI-NRQCD case are again almost negligible, while the maximized $f$-type contribution is smaller as compared to the corresponding one in the positive $x_F$ region and already in qualitative agreement with PHENIX data, leaving small room, if any, for constraining the gluon Sivers function. Notice however that this is also due to an approximate coincidence of the average $P_T$ of the PHENIX data with the region of crossing between negative and positive values of the CGI-NRQCD results. From this point of view, data at smaller or, even better, larger (around 3 GeV) average $P_T$ values would be very useful.

\begin{figure}[t]
\begin{center}
\includegraphics[width=0.42\textwidth]{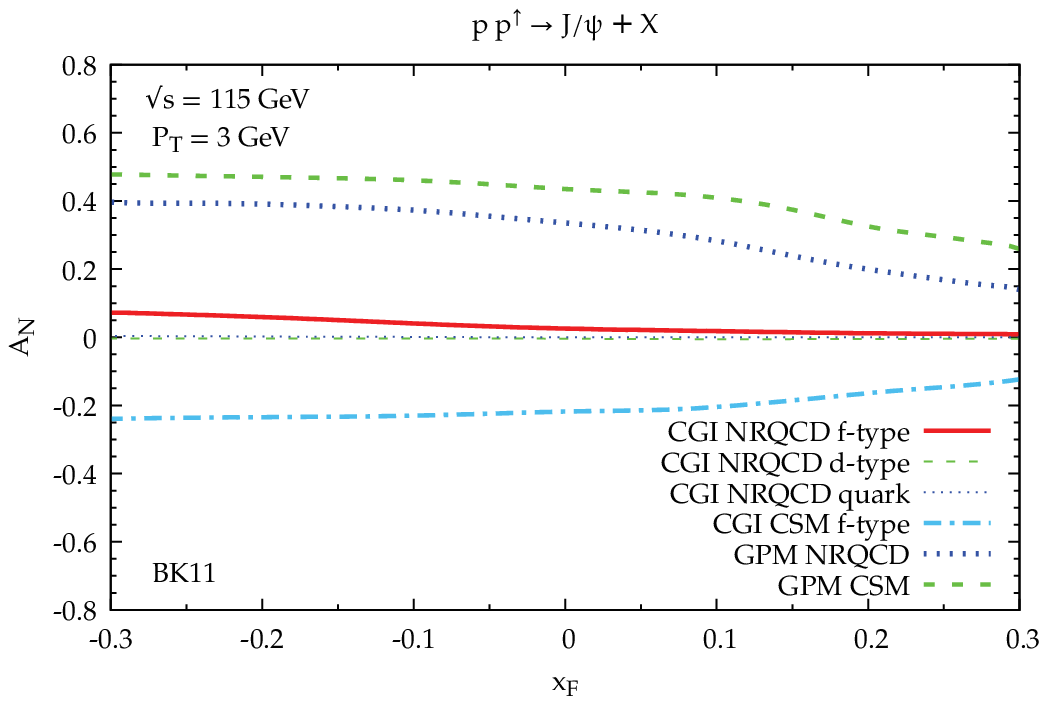}
\includegraphics[width=0.42\textwidth]{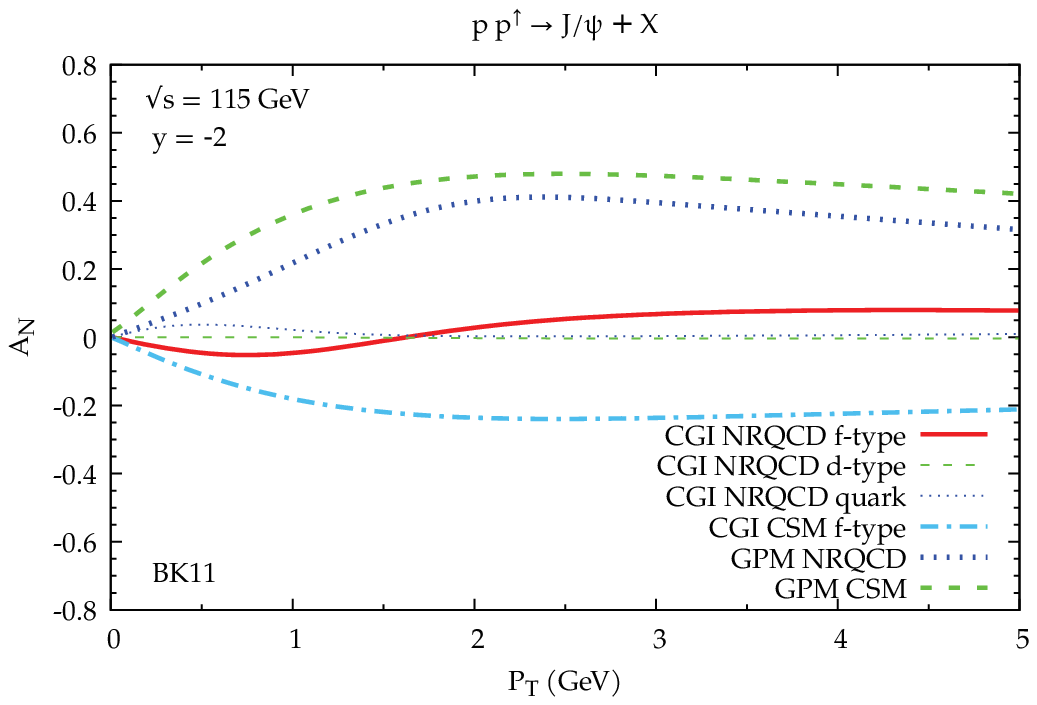}
\end{center}
\caption{Maximized values for $A_N$ for the process $p p^\uparrow \to J/\psi + X$ at $\sqrt s=115$ GeV and $P_T=3$ GeV as a function of $x_F$ (left panel) and at $y=-2$ as a function of $P_T$ (right panel), obtained adopting the CGI-GPM and GPM approaches, within the CS model and NRQCD (BK11 set). Notice that here negative rapidities correspond to the forward region for the polarized proton.
{\it Reprinted figure with permission from Ref.~\cite{DAlesio:2020eqo}, DOI:~https://doi.org/10.1103/PhysRevD.102.094011. Copyright (2020) by the American Physical Society}.}
\label{fig:ANptlhcb}
\end{figure}

RHIC is presently the only available high-energy proton-proton collider equipped with polarized beams. There are however other experimental setups that hopefully in the near future will be able to measure transverse SSAs in kinematical configurations suitable for inclusive quarkonium production and for other processes of fundamental phenomenological importance.   These will be extremely useful in testing the TMD approach, the validity of factorization and of universality and its modified but predictable versions.
We mention here the proposed plans for a (polarized) fixed target setup at the CERN Large Hadron Collider (LHC), the AFTER~\cite{Bjorken:2018} and LHCSpin~\cite{Aidala:2019pit} projects. An unpolarized fixed target has been already successfully inserted in the LHCb experiment for the measurement of nuclear cross sections of astrophysical interest, the SMOG experiment.
The LHCSpin proposal plans to continue and extend this programme by utilizing polarized Hydrogen and Deuterium targets. The AFTER project is a more general proposal considering and promoting all possible options (including the LHCSpin one) for performing spin physics at the LHC with a polarized fixed target.
It is therefore interesting to give estimates for $A_N(p p^\uparrow \to J/\psi + X)$ in kinematical configurations similar to those planned for the experimental setups proposed in the LHCSpin and AFTER projects.
To this end, in Fig.~\ref{fig:ANptlhcb} we show the maximized contributions to $A_N(p p^\uparrow \to J/\psi + X)$ at $\sqrt{s} = 115$ GeV as a function of $x_F$ at $P_T = 3$ GeV (left panel), and as a function of $P_T$ at fixed rapidity $y=-2$ (right panel), using the BK11 LDME set.
Notice that, in contrast with the results of Figs.~\ref{fig:allxF01}--\ref{fig:allpt165}, in this fixed-target setup negative values of $y$ and $x_F$ correspond to the forward region for the polarized proton.
These results are qualitatively similar to those for the RHIC setup and show that, while in the GPM model the constraining power of experimental data could be significative, in the CGI-NRQCD case already the maximized contributions are relatively tiny and in order to possibly costrain the $f$-type GSF very precise data would be required.
Another important, near-future experimental setup is the NICA-SPD pro\-ject at JINR in Dubna (see e.g.~\cite{Guskov:2019qqt}), that will study spin effects in polarized (both single and double) proton-proton (for $12 \leq \sqrt{s} \leq 27$ GeV) and (for the first time) deuteron-deuteron ($4 \leq \sqrt{s_{_{NN}}} \leq 14$ GeV) collisions. Charmonium production is one of the main subjects of the NICA-SPD physics programme. A phenomenological study within the GPM and CGI-GPM and the NRQCD approaches is currently in progress (see also~\cite{Karpishkov:2020brv}).

Before closing this section, let us briefly remind that, as mentioned before, data on SSAs for jet~\cite{Adamczyk:2012qj}, neutral pion~\cite{Adare:2013ekj} and muons (from $D$-meson decay)~\cite{Aidala:2017pum} production at mid-central rapidity and moderately large $P_T$ are also available from RHIC experiments.
In particular, pion SSA data are very accurate.
As compared to the $J/\psi$ production case, these processes involve in the final state TMD fragmentation functions that, like the LDMEs of NRQCD, need to be parametrized by comparison with data. Moreover, for meson production, other contributions in addition to the Sivers effect (namely, the Collins effect) can play a role.
A combined analysis of these results in the GPM and CGI-GPM approaches was performed~\cite{DAlesio:2017rzj,DAlesio:2018rnv}, showing that in the kinematical configurations considered the Sivers effect is in any case the dominant contribution. Based on this, and adopting (for the pion case) quark Sivers functions as extracted from SIDIS, the constraining power of these SSA data on the $f$- and $d$-type GSFs was investigated.
A first outcome is that while in neutral pion production the $f$-type GSF plays a dominant role as compared to the $d$-type one, the opposite is true in $D$-meson production.
A first exploratory study, in which the factors ${\cal N}^{(f,d)}_g(x_g)$, see Eq.~(\ref{eq:nq-coll}), were
kept constant (namely, independent of $x_g$) but allowed to vary with respect to the value $|{\cal N}^{(f,d)}_g(x_g)| = +1$ that saturates the positivity bound, was performed. It turns out that already in this simplified analysis the SSA data available put significant constraints on the indicative overall size of the GSFs in the CGI-GPM approach. More specifically, in order to get a reasonably good description of the $\pi^0$ and $D\to \mu$ SSA data we have to take $-0.15 \leq {\cal N}_g^{(d)} \leq + 0.15$ and correspondingly  $+0.05 \geq {\cal N}_g^{(f)} \geq -0.01$. In other words, both the $f$- and $d$-type GSFs are strongly constrained in size, and the $f$-type one even more effectively. As for the GPM scheme, the $D\to \mu$ SSA data seem not to constrain further the GSF as compared to the neutral pion data, which give already a very strong constraint~\cite{DAlesio:2018rnv}.

\begin{figure}[t]
\begin{center}
\includegraphics[width=0.45\textwidth]{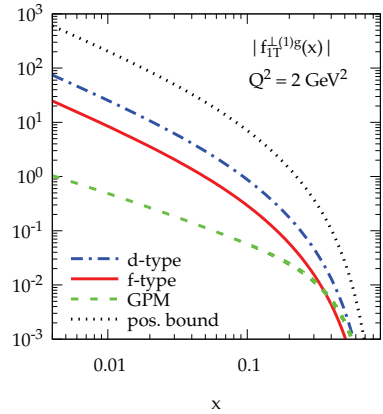}
\caption{Upper values for the first ${k}_\perp$-moments of the gluon Sivers functions in different approaches and scenarios at $Q^2 = 2$ GeV$^2$: GPM approach (green dashed line), CGI-GPM $d$-type (${\cal N}_g^{(d)}=0.15$, blue dot-dashed line) and $f$-type (${\cal N}_g^{(f)}=0.05$, red solid line). The positivity bound (black dotted line) is also shown. {\it Reprinted figure with permission from Ref.~\cite{DAlesio:2018rnv}, DOI:~https://doi.org/10.1103/PhysRevD.99.036013. Copyright (2019) by the American Physical Society}.}
\label{fig:1stmom}
\end{center}
\end{figure}

The results of this analysis can be summarized in terms of the extracted first ${k}_\perp$-moments of the TMD GSFs (see Fig.~\ref{fig:1stmom}):
\begin{equation}
\Delta^N \! f_{g/p^\uparrow}^{(1)}(x) = \int {\rm d}^2 \bm{k}_\perp \frac{k_\perp}{4 M_p} \Delta^N \! f_{g/p^\uparrow}(x,k_\perp) \equiv - f_{1T}^{\perp (1) g}(x) \, .
\label{siversm1}
\end{equation}
\section{Conclusions}
\label{sec:con}
In this contribution we discussed two main topics. Firstly, we presented a short overview of the so-called transverse momentum dependent approach and its application to polarization phenomena and the study of azimuthal and spin asymmetries in inclusive and semi-inclusive high-energy particle production. We kept the discussion at a very qualitative level, our aim being that of offering to non-specialized readers
a quick summary of the main ideas.
In the second, more technical part, we considered a current application of the TMD scheme, namely the GPM approach and its colour-gauge invariant extension, focusing on single spin asymmetries in inclusive quarkonium production in polarized proton-proton collisions.
Quarkonium production and decay mechanisms are of interest by themselves, see e.g.~the recent discovery of several new and unexpected resonances, or the still open problem of $J/\psi$ spin alignment, to quote just two hot topics.
The study of quarkonium production in (un)polarized processes can be of great help for learning about the poorly known TMD gluon distributions. As we have shortly illustrated, the study of these phenomena is a big challenge for QCD theoretical approaches, as far as concerns factorization theorems, the simultaneous employment of the TMD and NRQCD approaches, eventually in combination, at very high cm energies, with small-$x$ physics (a subject not covered here). At the same time, they offer a valuable and unique opportunity for shedding light on the physical mechanisms involved and the theoretical approaches developed for their study, deserving further efforts in this direction. As we have mentioned, besides the current rich programme on (un)polarized quarkonium production at RHIC, future promising opportunities will be offered by the proposed and planned quarkonium and spin physics road maps at the LHC (AFTER and LHCSpin proposals) and NICA. A very rich plan on single and double semi-inclusive quarkonium production is also planned for the future electron ion collider (EIC)~\cite{Accardi:2012qut}, which will offer a unique additional tool for learning about the TMD gluon distributions.

\begin{acknowledgements}
We are grateful to our colleagues C.~Flore, L.~Maxia, S.~Rajesh and P.~Taels for their collaboration in the quarkonium and spin physics research programme. This work is financially supported by Fondazione di Sardegna under the projects ``Quarkonium at LHC energies", project number F71I17000160002 (University of Cagliari), and
``Proton tomography at the LHC", project number F72F20000220007 (University of Cagliari), and by the European Union’s Horizon 2020 research and innovation programme under grant agreement N. 824093.
\end{acknowledgements}


%
%

\end{document}